\documentclass[sigconf,,nonacm]{acmart}

\AtBeginDocument{%
  \providecommand\BibTeX{{%
    \normalfont B\kern-0.5em{\scshape i\kern-0.25em b}\kern-0.8em\TeX}}}

\graphicspath{{./images/}} 

\usepackage{braket}
\usepackage{hyperref}
\usepackage{xcolor}
\usepackage{multirow}
\usepackage{multicol}
\usepackage{physics}
\begin{document}

\title{HyQBench: A Benchmark Suite for Hybrid CV-DV Quantum Computing}

\author{Shubdeep Mohapatra}
\affiliation{%
  \institution{North Carolina State University}
  \city{Raleigh}
  \state{North Carolina}
  \country{USA}}
\email{smohapa5@ncsu.edu}

\author{Yuan Liu}
\affiliation{%
  \institution{North Carolina State University}
  \city{Raleigh}
  \state{North Carolina}
  \country{USA}}
\email{q_yuanliu@ncsu.edu}

\author{Eddy Z. Zhang}
\affiliation{%
 \institution{Rutgers University}
 \city{Piscataway}
 \state{New Jersey}
 \country{USA}}
\email{eddy.zhengzhang@gmail.com}

\author{Huiyang Zhou}
\affiliation{%
  \institution{North Carolina State University}
  \city{Raleigh}
  \state{North Carolina}
  \country{USA}}
\email{hzhou@ncsu.edu}

\renewcommand{\shortauthors}{}
\begin{abstract}
Hybrid continuous-variable (CV)–discrete-variable (DV) quantum systems present a promising direction for quantum computing by combining the high-dimensional encoding capabilities of qumodes with the control offered by DV qubits on the coupled qumodes. There have been exciting recent progresses on hybrid CV-DV quantum computing, including variational algorithms, error correction, compiler-level optimizations for Hamiltonian simulation, etc. However, there is a lack of a standardized CV-DV benchmark suite for assessing various emerging hardware platforms and evaluating software optimizations on hybrid CV-DV circuits.

In this work, we introduce a simulation and benchmarking framework for hybrid CV-DV circuits, implemented using \textit{Bosonic Qiskit}\cite{bosonic-qiskit}—a tool specifically designed to model CV-DV systems, along with QuTip\cite{qutip} for functional correctness verification. We construct and characterize representative CV-DV benchmarks, including cat state generation, GKP state generation, CV-DV state transfers, hybrid quantum Fourier transform, variational quantum algorithms, Hamiltonian simulation, and Shor's algorithm.

To assess circuit complexity and scalability, we define a feature map organized into two categories: general features (e.g., qubit/qumode count, gate counts) and CV-DV-specific features (e.g., Wigner negativity, energy, truncation cost). These metrics enable evaluation of both classical simulability and hardware resource requirements.

Our results, including one benchmark on real hardware, demonstrate that hybrid CV-DV architectures are not only viable but well-suited for a range of computational tasks, from optimization to Hamiltonian simulation. This framework lays the groundwork for systematic evaluation and future development of hybrid quantum systems.
\end{abstract}
\maketitle

\section{Introduction}

Quantum computing aims to harness the principles of quantum mechanics to solve problems that are intractable or inefficient on classical computers \cite{qc_tb}. By exploiting quantum phenomena such as superposition and entanglement, certain algorithms—most notably Shor’s algorithm for integer factorization—offer theoretical speedups over their classical counterparts. However, much of the existing research has focused predominantly on discrete-variable (DV) systems, which are typically composed of two-level quantum units realized using platforms such as superconducting transmons \cite{transmon1,transmon2}  electronic spins \cite{spin_qc} or trapped ions \cite{trapped-ion-chrismonroe,trapped-ion-chrismonroe2,trapped-ion-scaling}
 and neutral atoms\cite{neutral_atoms_demo,neutral-atoms}.

An alternative paradigm involves using bosonic modes as computational resources\cite{cv-qc}. Bosons exist, in principle, in an infinite-dimensional Hilbert space, enabling them to be naturally described as continuous-variable (CV) systems. CV architectures facilitate efficient simulation of bosonic dynamics and provide higher-dimensional encoding spaces that can be exploited for optimization and quantum information processing tasks. However, CV-only systems face significant challenges in implementing non-Gaussian operations and achieving fault tolerance. This has motivated the recent development of hybrid architectures that combine the discrete control and universality of qubits with the higher-dimensional Hilbert space of bosonic modes. Such hybrid CV-DV systems represent a promising path toward scalable and versatile quantum computing platforms. 

Despite growing interest, the literature on hybrid CV-DV systems remains relatively sparse \cite{cvdv-literature,bosonic-isa,genesis}. As a foundational step, it is important to assess their viability across metrics such as scalability, performance, and fidelity. To this end, developing a dedicated benchmarking suite is essential for evaluating how hybrid CV-DV systems perform on representative tasks such as Hamiltonian simulation, quantum Fourier transform (QFT), and optimization problems. Unlike the DV domain, which benefits from well-established benchmarking suites like SuperMarQ\cite{tomesh2022supermarq} and QASMBench\cite{QASMBench}, there is currently no comprehensive benchmarking framework tailored specifically for hybrid CV-DV architectures.

Some simulation tools, such as Bosonic Qiskit\cite{bosonic-qiskit}, have begun to address this gap by representing bosonic modes with multiple qubits; for example, using \( n \) qubits to encode \( 2^n \) states, thereby imposing a \textit{cutoff} on the Hilbert space dimension of a qumode. However, a unified benchmarking framework that captures the hybrid nature and unique characteristics of CV-DV systems remains an open research need.

In this paper, we present HyQBench, a comprehensive benchmark suite implemented using Bosonic Qiskit and QuTip: gate-level circuit model in Bosonic Qiskit and matrix-based computation in QuTip for functional correctness verification. HyQBench covers a wide range of circuits, including QFT, variational quantum algorithms, and other application-specific benchmarks. We also include primitive circuits such as GKP state generation, which serves as input state preparation for more complex benchmarks. Additionally, we propose metrics to characterize these benchmarks, including general resource metrics, such as gate counts and qumode counts, as well as CV-DV specific features, like Wigner negativity, which provides insight into the non-classicality of a CV state and indicates the classical simulation complexity of the system.

The key motivations for the development of \textbf{HyQBench} include showcasing the advantage of hybrid CV-DV systems and democratizing this emerging quantum computing paradigm. Solving the Jaynes-Cummings-Hubbard (JCH) model, which describes interactions between two-level systems(TLS) and bosonic oscillators using DV-only or CV-only, is inefficient. DV systems require large qubit registers to encode oscillator states and the interactions between the TLS and the oscillator, while CV systems lack native TLS interactions. Hybrid CV–DV architectures bridge this gap by representing bosonic degrees of freedom as qumodes and TLS components as qubits, enabling direct mapping of the Hamiltonians that are inherently hybrid.
This mapping offers substantial resource advantages. For instance, simulating a 3-site JCH Hamiltonian (each cavity truncated at four photons) in a DV-only encoding synthesized with mat2qubit\cite{mat2qubit} and Qiskit requires \textbf{9 qubits}, \textbf{393 CNOT}, and \textbf{265 U3} gates for a single application of the full Hamiltonian evolution $e^{-iHt}$. In contrast, the hybrid CV–DV realization performs one Trotter step using \textbf{3 CV modes}, \textbf{3 qubits}, \textbf{3 U3 gates}, \textbf{3 Phase Space Rotation gates}, \textbf{3 Jaynes Cummings} and \textbf{2 beamsplitter gates}. 
This JCH Hamiltonian simulation is just one benchmark included in HyQBench, which demonstrates that hybrid CV–DV systems are a resource-efficient framework for multiple application domains.

The benchmark suite introduced in this work is designed not only for performance evaluation but also as a calibration framework currently used by the QSCOUT\cite{qscout} team. At the time of writing, the cat-state preparation circuit is the first benchmark run on hardware and serves as the primary experimental validation reported here, while the remaining benchmarks constitute the ongoing calibration and validation roadmap and will be executed as device calibration progresses.

Overall, the paper makes the following contributions:
\begin{itemize}
  \item An open-source comprehensive \href{https://github.com/shubdeepmohapatra01/HyQBench/tree/main}{\textbf{benchmark suite}} to showcase the usage of hybrid CV-DV quantum circuits and to assess hardware and/or software development on this emerging architecture.
  \item A set of metrics to characterize hybrid CV-DV circuits.
  \item An analysis to show the advantages of hybrid CV-DV systems over DV-only and CV-only systems for the benchmarks.
  \item A noise analysis on the benchmarks. 
  \item The experimental results of one benchmark on real quantum hardware.
\end{itemize}

\section{Background}

\subsection{Discrete Variable (DV) Systems}
In DV systems, the state information is encoded within a finite-dimensional Hilbert space. The fundamental unit of a DV system is a qubit, which is a two-level system. A qubit can exist in either state $\ket{0}$ or $\ket{1}$, or in a superposition of both. A general state of a DV system can be represented as:
\[
\ket{\psi} = \alpha \ket{0} + \beta \ket{1}, \quad \text{where } \alpha, \beta \in \mathbb{C},\ |\alpha|^2 + |\beta|^2 = 1
\]

Manipulation of the qubit state is achieved by performing unitary operations $U$ on the qubit. Any single-qubit operation can be represented in terms of the three Pauli matrices:
\[
X = \begin{pmatrix}
0 & 1 \\
1 & 0
\end{pmatrix}, \quad
Y = \begin{pmatrix}
0 & -i \\
i & 0
\end{pmatrix}, \quad
Z = \begin{pmatrix}
1 & 0 \\
0 & -1
\end{pmatrix}
\]

To extend this into the multi-qubit regime, entangling gates that act on multiple qubits are needed. One such gate is the CNOT gate.

A set of gates like $S = \{X(\theta), Z(\theta), \mathrm{CNOT}\}$, forms a universal gate set meaning that any unitary operation on an $n$-qubit system can be approximated to arbitrary precision using finite sequences of gates from that set. Here, $X(\theta) = e^{-i\theta X/2}$ and $Z(\theta) = e^{-i\theta Z/2}$ are single-qubit rotation gates about the $x$ and $z$-axes of the Bloch sphere, respectively. There are also discrete universal gate sets, the minimal gate set $S' = \{H, T, \mathrm{CNOT}\}$. The Solovay-Kitaev theorem provides bounds on the number of gates needed from the set $S'$ to approximate any unitary operation up to an error of $\epsilon$\cite{solovaykitaevalgorithm}.

\subsection{Continuous Variable (CV) Systems}
Continuous-variable (CV) quantum systems encode information in observables with continuous spectra, in contrast to discrete-variable (DV) qubits. A single CV mode (qumode) is modeled as a quantum harmonic oscillator with Hamiltonian
\[
\hat{H}=\frac{1}{2}(\hat{p}^2+\hat{x}^2)=\hat{a}^\dagger\hat{a}+\frac{1}{2},
\]
in units where $\hbar\omega=1$. The canonical operators satisfy $[\hat{x},\hat{p}]=i$, and the ladder operators
\[
\hat{a}=\frac{1}{\sqrt{2}}(\hat{x}+i\hat{p}), \quad 
\hat{a}^\dagger=\frac{1}{\sqrt{2}}(\hat{x}-i\hat{p})
\]
obey $[\hat{a},\hat{a}^\dagger]=1$.

\subsubsection{Fock Basis and States}
Eigenstates of $\hat{n}=\hat{a}^\dagger\hat{a}$ form the number basis $\{\ket{n}\}$ with
\[
\hat{n}\ket{n}=n\ket{n}, \qquad 
\hat{a}\ket{n}=\sqrt{n}\ket{n-1}, \quad 
\hat{a}^\dagger\ket{n}=\sqrt{n+1}\ket{n+1}.
\]
Any state can be expanded as $\ket{\psi}=\sum_n c_n\ket{n}$.

\subsubsection{Representation of CV States}
The Fock basis is one of several equivalent ways to describe a CV quantum state. Two other commonly used representations are:

\begin{itemize}

    \item \textbf{Wavefunction Representation (Position or Momentum):}
    States may also be described in the position or momentum basis,
    \[
    \psi(x)=\braket{x}{\psi}, \qquad \phi(p)=\braket{p}{\psi},
    \]
    which are related by Fourier transform.
    \item \textbf{Phase-Space Representation (Wigner Function):}
    The Wigner function provides a quasiprobability distribution,
    \[
    W(x,p)=\frac{1}{\pi}\int dy\,e^{2ipy}\bra{x-y}\rho\ket{x+y},
    \]
    It is real-valued but may be negative, signaling non-Gaussianity (Gaussian states have non-negative Wigner functions). This representation is widely used in CV quantum optics and simulation.
    
\end{itemize}

\subsubsection{CV Gates}
CV states are transformed by CV gates, which fall into two categories: Gaussian and non-Gaussian.

\paragraph{Gaussian Gates:}  
These gates are generated by Hamiltonians that are at most quadratic in the canonical operators $\hat{x}$ and $\hat{p}$ (or equivalently in $\hat{a}$ and $\hat{a}^\dagger$). Gaussian gates can be efficiently simulated classically \cite{gaussian-gates}.
Common Gaussian gates include Displacement, Squeezing, and Rotation Gates. Additionally, the Beamsplitter gate is an example of a multi-qumode Gaussian gate.

\paragraph{Non-Gaussian Gates:}  
Non-Gaussian gates introduce nonlinearities and map Gaussian states to non-Gaussian ones. These gates are generated by Hamiltonians involving higher-than-quadratic terms in $\hat{x}$ and $\hat{p}$. 

Non-Gaussian gates are more resource-intensive to implement and simulate, but are necessary for universal quantum computation within the qumode. A parallel can be drawn from the DV scenario in that Gaussian gates, similar to Clifford Gates, are efficiently simulatable classically, in contrast to non-Clifford and non-Gaussian gates.

\begin{figure}
\centering
  \includegraphics[scale=0.4]{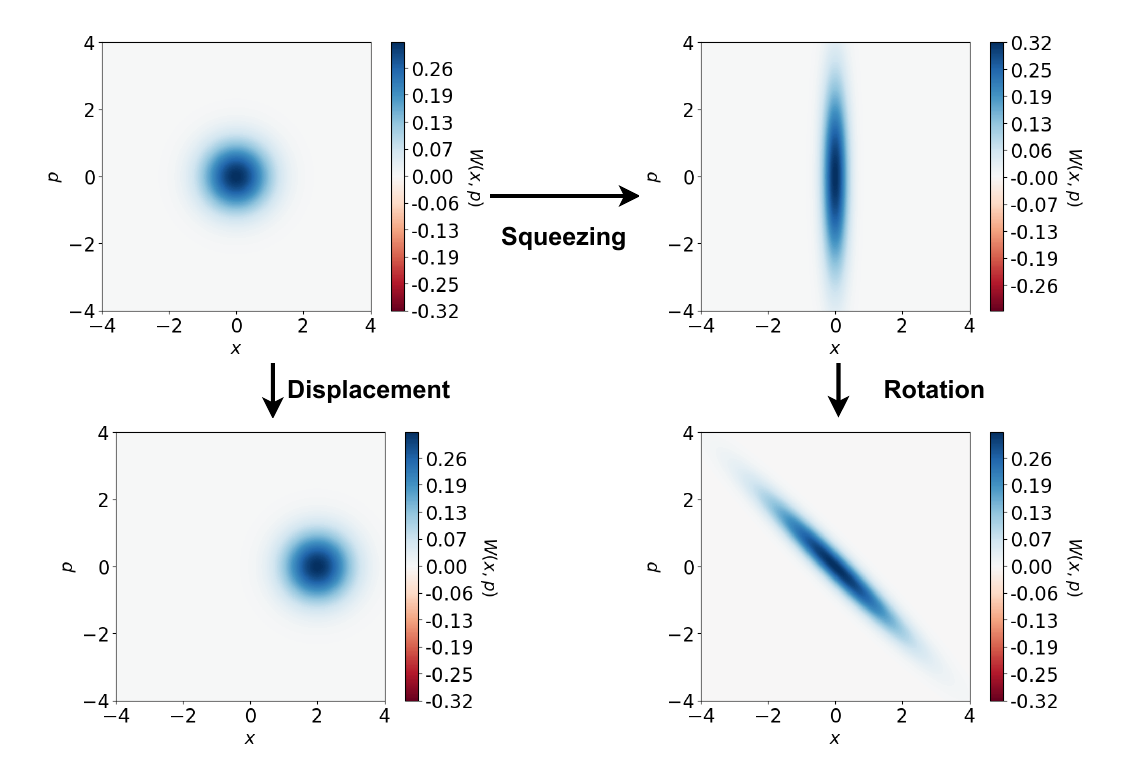}
  \caption{Effect of some basic CV gates in the phase space.}
  \label{fig:gates}
\end{figure}

Fig.~\ref{fig:gates} illustrates the action of some basic CV gates. The initial state is on the top left and is the phase space representation of the vacuum State $\ket{0}$. In the phase space representation, the x-axis corresponds to the position value, and the y-axis corresponds to the momentum value. The action of single mode squeezing gate squeezes the vacuum state in the position or momentum direction. The top right of Fig~\ref{fig:gates} shows the vacuum state being squeezed in the position axis. Doing so increases the certainty of finding the oscillator at $x=0$ but decreases the certainty in the momentum axis. The displacement gate displaces a state by the amount specified. The bottom left of Fig~\ref{fig:gates} shows an example of the vacuum state being shifted by an amount of 2 in the position axis. The rotation gate acting on the squeezed vacuum state, as shown in the bottom right, just rotates the state in the phase space (by $\frac{\pi}{4}$). %In this case, the squeezed vacuum state was rotated by $\frac{\pi}{4}$.

\subsection{Hybrid CV-DV Architecture}

Hybrid quantum systems have previously been used on a small scale for tasks such as error correction in superconducting platforms\cite{cvdv-qec,cvdv-qec1}. In superconducting quantum technology, a dispersive interaction naturally arises between the microwave resonator (a CV system) and the qubit. Each resonator is typically coupled to its nearest neighbor, enabling quantum information to be efficiently manipulated and stored through qubit-resonator and resonator-resonator operations. Other technologies such as trapped ions\cite{trapped-ion-chrismonroe,trapped-ion-chrismonroe2} and neutral atoms\cite{neutral-atom-cvdv}, while predominantly focused on qubits, have recently been explored and utilized to realize bosonic modes for hybrid computations.

% (???)

\subsubsection{Hybrid CV-DV Gates}

Hybrid CV--DV gates are typically qubit-controlled versions of standard CV operations. Two key examples used throughout our benchmarks are:

\begin{itemize}
    \item \textbf{Conditional Displacement:} 
    \[
        D_c(\alpha,\beta)=\dyad{0}\otimes D(\alpha)+\dyad{1}\otimes D(\beta).
    \]
    A common special case is the \emph{symmetric conditional displacement},
    \[
    CD(\alpha)=D_c(+\alpha,-\alpha)=\exp\!\left[\sigma_z(\alpha \hat a^\dagger-\alpha^*\hat a)\right],
    \]
    which shifts the qumode by $+\alpha$ for qubit state $\ket{0}$ and by $-\alpha$ for $\ket{1}$. Real $\alpha$ produces a position shift; imaginary $\alpha$ produces a momentum shift.

    \item \textbf{Conditional Rotation:} $CR(\theta)=e^{-i\frac{\theta}{2}\sigma_z \hat{a}^\dagger \hat{a}}, $
    implementing a qubit-dependent phase rotation of the qumode, naturally arising from the dispersive interaction in hybrid systems.
\end{itemize}

\subsection{Advantages of CV-DV Systems}
There are several advantages by coupling a CV system with a DV system. \textbf{First}, the higher degrees of freedom in a qumode provide more computational "space." Note that for classical simulation, we truncate the qumode to a finite number of Fock Levels.
\textbf{Second}, systems containing qumodes are often more efficient at simulating systems involving bosons. Qubit-only systems are well-suited for simulating spin dynamics; however, simulating fermionic and bosonic systems introduces significant overhead. The reason is that although a Fock cutoff of $N$ can, in principle, be achieved using $\log_2 N$ qubits, accurately implementing the annihilation ($\hat{a}$) and creation ($\hat{a}^\dagger$) operators in a qubit-only system requires a deep and complex gate sequence. In contrast, these operators are natively supported and easily applied in CV systems, offering a more resource-efficient implementation. \textbf{Third}, coupling qubits to CV systems offers a key benefit, since non-Gaussian gates, which are required for universal computation, can be implemented more easily and efficiently using a qubit to control the qumode \cite{cv-dv-adv2}, compared to CV-only systems. \textbf{Fourth}, compared to CV-only systems, the resource requirements for fault-tolerant operations are much lower in the CV-DV systems \cite{cv-dv-adv1}.

\subsection{Related Work}
Benchmarking in discrete-variable (DV) systems has been extensively studied, with a focus on evaluating different aspects of quantum circuits. One of the most widely adopted approaches is Randomized Benchmarking \cite{RB-2008,RBS-2011,RBS-2012,RBS-2013,RBS-2014}, which applies sequences of randomly selected quantum gates to estimate average gate fidelity. This technique is particularly effective at capturing coherent errors. % that may not be detected by traditional methods. 

Beyond gate-level metrics, several application-oriented benchmarking suites have emerged. SupermarQ~\cite{tomesh2022supermarq} includes a diverse set of workloads, such as Variational Quantum Eigensolvers (VQE), Quantum Approximate Optimization Algorithm (QAOA), and GHZ state preparation, and evaluates metrics like parallelism, critical depth, and connectivity. QASMBench~\cite{QASMBench} offers a collection of quantum circuits across various scales (small, medium, and large) to assess performance on different hardware backends. 
MQTBench \cite{MQTBench} also provides a comprehensive benchmarking suite catering to different level of abstractions.

For continuous-variable (CV) systems, simulation frameworks such as Bosonic Qiskit \cite{bosonic-qiskit} and Xanadu's Strawberry Fields\cite{strawberryfields} and PennyLane\cite{pennylane} have been developed to enable modeling of qumode-based circuits. 

Recent advances in hybrid CV-DV architectures have demonstrated the potential of combining the strengths of both paradigms. ~\cite{bosonic-isa} introduced a hybrid instruction set architecture (ISA) and proposed a full-stack CV-DV system capable of supporting both qubit and qumode operations. Application-specific studies have also been explored—for instance, VQE optimization for ground state energy calculation \cite{cvdv-vqe1}, CV-DV QML \cite{cvdv-qml} and a CV-DV implementation of the Quantum Fourier Transform \cite{ad/da-cvdv}. Additionally, \cite{genesis} proposed a hybrid CV-DV compiler called \textbf{Genesis}, which decomposes bosonic-fermionic Hamiltonians into native gate sets tailored to hybrid hardware platforms.

Despite these promising developments, there is currently no established benchmarking suite for hybrid CV-DV systems that parallels the comprehensiveness of DV benchmarking frameworks. Specifically, there is a lack of standardized tools that consolidate both primitive gates and application-specific workloads within a single benchmarking environment.

In this work, we address this gap by proposing HyQBench, a comprehensive benchmarking suite for hybrid CV-DV quantum systems. HyQBench spans a broad range of circuit templates, application types, and complexity levels, enabling systematic evaluation of hybrid architectures in terms of resource cost, non-Gaussianity, and truncation overhead.

\subsection{Early Demonstration of CV-DV Hardware}
Recently, there have been advancements in real hardware for hybrid CV-DV systems. One example is IQM Quantum computers, which have implemented a Hybrid CV-DV Move gate in their machine, IQM Crystal\cite{iqm-resonance}. This machine has a star topology, where all qubits connect to a central resonator.  
Although currently there is no support for direct measurement of the resonator, its state can be verified by comparing the qubit states since the probabilities are a function of the Fock number $n$\cite{iqm-resonance}. 

 QSCOUT \cite{qscout}, a trapped-ion system from Sandia National Labs uses the motional modes of the ions as CV modes and the ions as a DV resource. Furthermore, the QSCOUT team provides experimental support for running one of our benchmarks as presented in Section IX. 

\section{HyQBench Benchmarks}
\begin{table}[t]
\caption{Summary of Benchmarks}
\label{tab:benchmark_summary}
\centering
\renewcommand{\arraystretch}{1.2}
\begin{tabular}{|l|p{5.2cm}|}
\hline
\textbf{Benchmark} & \textbf{Description} \\ \hline
State Transfer & Transfers a qumode state to DV qubits or vice-versa. \\ \hline
CAT State Protocol & Deterministically generates an even cat state. \\ \hline
GKP State Protocol & Builds an approximate GKP state by iteratively applying the CAT state protocol. \\ \hline
CV-DV QFT & Performs DV QFT by using a qumode. \\ \hline
CV-QAOA & Solves continuous minimization problems using a qumode. \\ \hline
CV-DV VQE & Solves a binary knapsack problem using a hybrid CV-DV variational ansatz. \\ \hline
JCH Simulation & Simulates the Jaynes–Cummings–Hubbard model using the CV-DV system. \\ \hline
Shor's Algorithm & Factors integers using a CV-DV implementation of Shor’s algorithm. \\ \hline
\end{tabular}
\end{table}

Table~\ref{tab:benchmark_summary} provides an overview of the eight benchmarks studied in this work, summarizing their primary objectives. Our benchmarks are designed to be representative and comprehensive, encompassing three hierarchical levels. At the primitive level, they include fundamental operations such as state preparation and state transfer. The algorithmic level benchmarks, which include QFT, CV-DV VQE, and CV-QAOA, capture mid-level circuit complexity. Finally, at the application level, benchmarks like Shor’s algorithm and Hamiltonian Simulation showcase higher-level computational problems. %More benchmarks can be incorporated when new CV-DV applications are developed. 

\subsection{State Transfer Circuit}
\begin{figure}
\centering
  \includegraphics[scale=0.38]{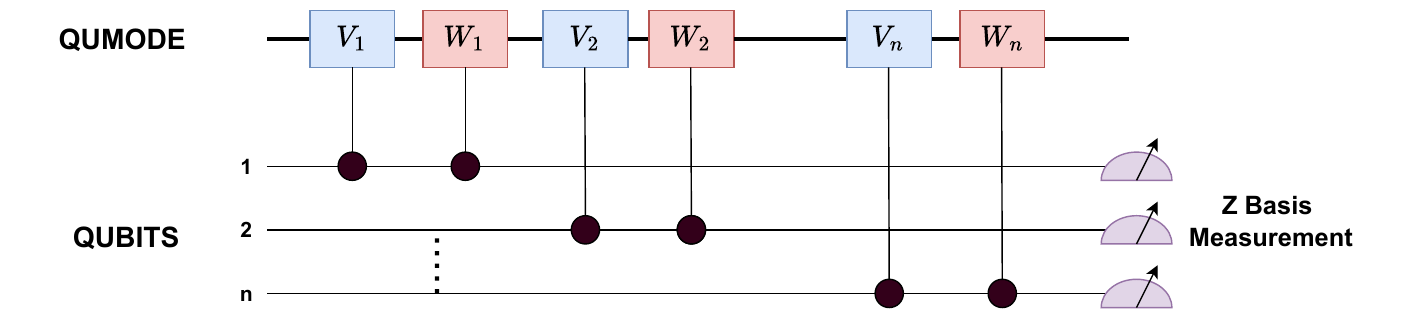}
  \caption{The CV to DV State Transfer Circuit. Note that the gates have to be in reverse order for DV to CV state transfer. The gates $V_j$ and $W_j$ are hybrid qumode-qubit gates. We measure all qubits to extract the CV state}
  \label{fig:state_transfer_circuit}
\end{figure}
The state transfer circuit provides a protocol to transfer quantum information between a qumode and $N$ qubits. Based on the scheme proposed in \cite{non-abelian-statetransfer}, the circuit achieves bidirectional transfer: $U^\dagger(\Delta, N)$ maps a CV state to an $N$-qubit state, while $U(\Delta, N)$ performs the reverse transfer. The protocol uses a sequence of conditional displacement gates that couple the qumode to each qubit, enabling mapping of quantum states between the CV and DV subsystems. The parameter $\Delta$ controls the spacing of the displacements and is tuned according to $N$.

The unitary $U^\dagger(\Delta, N)$ is defined as
\begin{equation}
    U^\dagger(\Delta, N) = \prod_{j=N}^{1} W_j V_j = W_N V_N \cdots W_1 V_1,
\end{equation}
where
\begin{align}
    V_j &= e^{i \frac{\pi}{\Delta2^j}  \hat{x} \hat{\sigma}_y^{(j)}}, 
    W_j &=
    \begin{cases}
        e^{i \frac{\Delta}{2^{j-1}} \hat{p} \hat{\sigma}_x^{(j)}}, & j < N, \\
        e^{-i \frac{\Delta}{2^{j-1}} \hat{p} \hat{\sigma}_x^{(j)}}, & j = N.
    \end{cases}
\end{align}
Here, $V_j$ applies a momentum displacement on the qumode conditioned on the $j^{\text{th}}$ qubit via $\hat{x}$, while $W_j$ applies a position displacement conditioned on the qubit via $\hat{p}$, with the sign determined by $j$.

Fig~\ref{fig:state_transfer_circuit} shows the circuit implementation of this protocol.

\subsection{Deterministic Cat State Protocol }
\begin{figure}
\centering
  \includegraphics[scale=0.25]{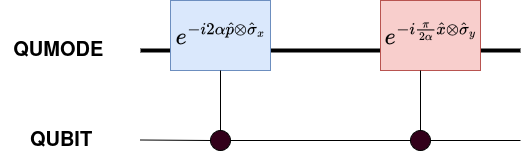}
  \caption{High-level circuit representation of the cat state protocol.}
  \label{fig:cat_circuit}
\end{figure}
\begin{figure}
\centering
  \includegraphics[scale=0.25]{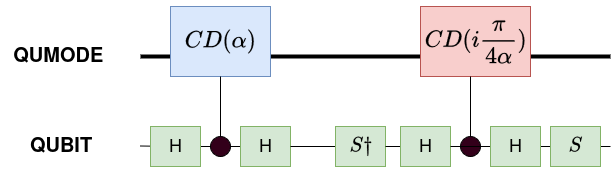}
  \caption{Decomposing the cat state protocol circuit into single-qubit gates and the conditional displacement gates.}
  \label{fig:cat_circuit_bq}
\end{figure}

The deterministic cat state protocol generates an even cat state, which is a superposition of two coherent states $\ket{\alpha}$ and $\ket{-\alpha}$, without requiring post-selection on an ancilla qubit measurement\cite{bosonic-isa}. Starting from the initial state $\ket{0}_{\text{osc}} \otimes \ket{0}$, the protocol consists of two main steps. First, a conditional displacement $e^{-i 2\alpha\, \hat{p} \otimes \hat{\sigma}_x}$ entangles the qubit and qumode, producing a superposition of $\ket{\alpha}$ and $\ket{-\alpha}$ correlated with the qubit states $\ket{-}$ and $\ket{+}$, respectively. Second, a disentangling operation $e^{-i \frac{\pi}{2\alpha} \hat{x} \otimes \hat{\sigma}_y}$ performs a qumode-controlled rotation of the qubit, leaving the qumode in an even cat state while separating it from the qubit. Both operations are realizable using conditional displacement gates available in the Phase-Space ISA, with appropriate basis changes on the qubit. The fidelity of the generated cat state w.r.t. the ideal one increases with the value of $\alpha$.

Fig~\ref{fig:cat_circuit} shows a high level representation of the Cat State protocol discussed above. Note that the first gate can be seen as a Conditional Displacement gate with real displacement and the second gate as a Conditional Displacement gate with imaginary displacement. Fig~\ref{fig:cat_circuit_bq} shows this decomposed circuit.

\subsection{GKP State Protocol}
\begin{figure}
\centering
  \includegraphics[scale=0.25]{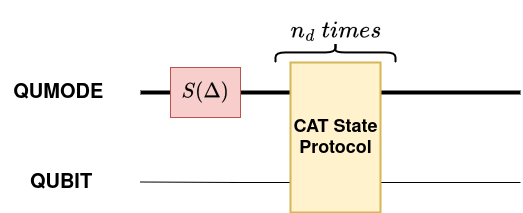}
  \caption{The GKP state generation circuit. The Cat State Protocol is repeated for $n_d$ times, and $S$ is the single-mode squeezing gate with the squeezing parameter $\Delta$}.
  \label{fig:gkp_circuit}
\end{figure}
The Gottesman-Kitaev-Preskill (GKP) state protocol generates approximate GKP states, which encode logical qubits into the infinite-dimensional Hilbert space of a qumode using a grid-like structure in phase space. As proposed in \cite{bosonic-isa}, the protocol leverages repeated applications of the deterministic cat state generation scheme, where successive conditional displacements create multiple evenly spaced coherent components. By fixing the displacement amplitude to $\alpha = \sqrt{\pi}$ and starting from a squeezed vacuum state, the resulting state approximates the ideal GKP logical states $\ket{0}_L$. The fidelity of the prepared state improves with increased squeezing of the initial state. As with the cat state protocol, the conditional displacement gate, natively available in the Phase-Space ISA, forms the key operation in this scheme.

Fig~\ref{fig:gkp_circuit} shows the circuit description for this protocol, which makes use of the cat state protocol in Fig~\ref{fig:cat_circuit_bq} here.

\subsection{CV-DV QFT}
\begin{figure}
\centering
  \includegraphics[scale=0.2]{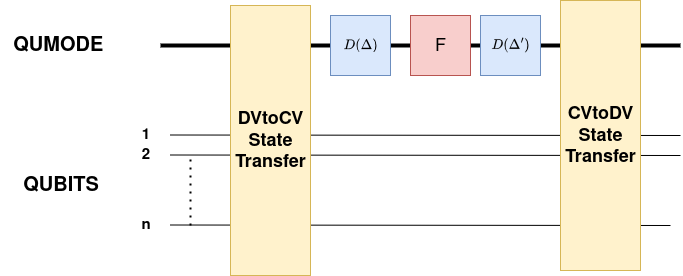}
  \caption{The CV-DV QFT circuit. It uses the state transfer protocol. The two free parameters are for the Displacement gates (D) $\Delta$ and $\Delta'$. Measure all qubits at the end.}
  \label{fig:qft_circuit}
\end{figure}
The CV-DV Quantum Fourier Transform (QFT) protocol exploits the property that a rotation of angle $\pi/2$ in the phase space acts as a Fourier gate, swapping the position and momentum operators up to a sign. As proposed in \cite{ad/da-cvdv}, the protocol performs the QFT on an $n$-qubit DV state in three steps: (i) transferring the DV state to a qumode using the DV-to-CV state transfer protocol, (ii) enacting free evolution under the Fourier gate on the qumode, and (iii) transferring the transformed state back to the $n$-qubit register using CV-to-DV state transfer. The fidelity of this hybrid QFT with the exact DV QFT improves when additional ancilla qubits initialized in $\ket{+}$ are appended to the input state, ensuring periodicity. The protocol relies on conditional displacement gates natively available in the Phase-Space ISA, while the F gate
is a virtual operation enacting the free evolution of the qumode.
Fig~\ref{fig:qft_circuit} shows the circuit representation of the CV-DV QFT protocol. $\Delta$ and $\Delta'$ are chosen such that the oscillator state is symmetric around $x=0$.

\subsection{CV-DV VQE}
The hybrid CV-DV variational quantum eigensolver (VQE) leverages the higher-dimensional Hilbert space of bosonic modes and constructs a universal ansatz for solving combinatorial optimization problems. As proposed in \cite{cvdv-vqe}, the ansatz uses single-qubit rotation gates together with Echoed Controlled Displacement (ECD) gates, which couple a qubit to a qumode via conditional displacements as realized on superconducting platform. This framework enables efficient encoding of binary variables: a qubit–qumode pair represents item selection variables, while an additional qumode encodes auxiliary slack variables. The Binary Knapsack Problem (BKP) is formulated in a QUBO framework, and the variational parameters of the gates are optimized to prepare a quantum state encoding the optimal bitstring. By exploiting the Fock cutoffs of the qumodes, this approach represents any number of binary variables with two qumodes and one qubit. The parameters of the ECD gates \((\beta)\) and the single-qubit rotation gates \((\theta,\phi)\) are optimized using classical optimizers within the VQE loop. In this benchmark, the photon number occupancy of the qumodes and the qubit state is measured.

\begin{figure}
\centering
  \includegraphics[scale=0.2]{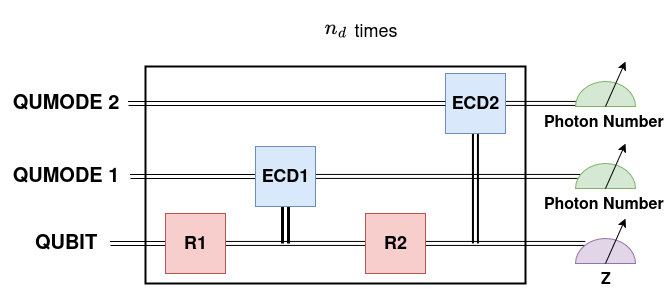}
  \caption{The CV-DV VQE ansatz (one layer). Note that the number of gates remains constant in each layer irrespective of problem size. The ECD gate can be decomposed into the controlled displacement gate using the rule $ECD = X.CD$}
  \label{fig:vqe-circuit}
\end{figure}

The overall circuit structure is illustrated in Fig.~\ref{fig:vqe-circuit}, and can be extended to arbitrary depth \( n_d \) as needed.

\subsection{CV QAOA}
\begin{figure}
\centering
  \includegraphics[scale=0.25]{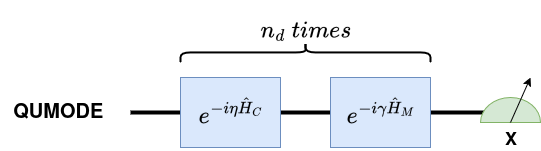}
  \caption{The CV QAOA circuit. The cost function unitary and mixer unitary are applied alternatively for a depth of $n_d$. These unitaries need be decomposed into hardware native gates for practical applications on the hardware.}
  \label{fig:qaoa_circuit}
\end{figure}
The continuous-variable Quantum Approximate Optimization Algorithm (CV-QAOA) extends the QAOA framework to qumodes, enabling optimization over continuous domains \cite{cv-qaoa}. The algorithm alternates between cost and mixer unitaries, where the cost Hamiltonian $\hat{H}_C = f(\hat{\mathbf{x}})$ encodes the objective function and the mixer Hamiltonian $\hat{H}_M = \frac{1}{2}\hat{p}^2$ causes position updates similar to gradient descent. Starting from a squeezed vacuum state in the momentum basis, the variational parameters $(\boldsymbol{\eta}, \boldsymbol{\gamma})$ are optimized classically to minimize the cost function. After applying $p$ QAOA layers, measuring the qumode in the $\hat{x}$-basis yields a candidate solution that approximates the minimum of $f(\hat{\mathbf{x}})$. This approach leverages the continuous nature of CV systems to efficiently perform variational optimization in continuous spaces.
Fig~\ref{fig:qaoa_circuit} shows the high level circuit for the CV QAOA algorithm. Note that, the cost and mixer unitaries have to be decomposed to hardware native gates.

\subsection{Jaynes-Cummings-Hubbard Simulation}

\begin{figure}
\centering
  \includegraphics[scale=0.3]{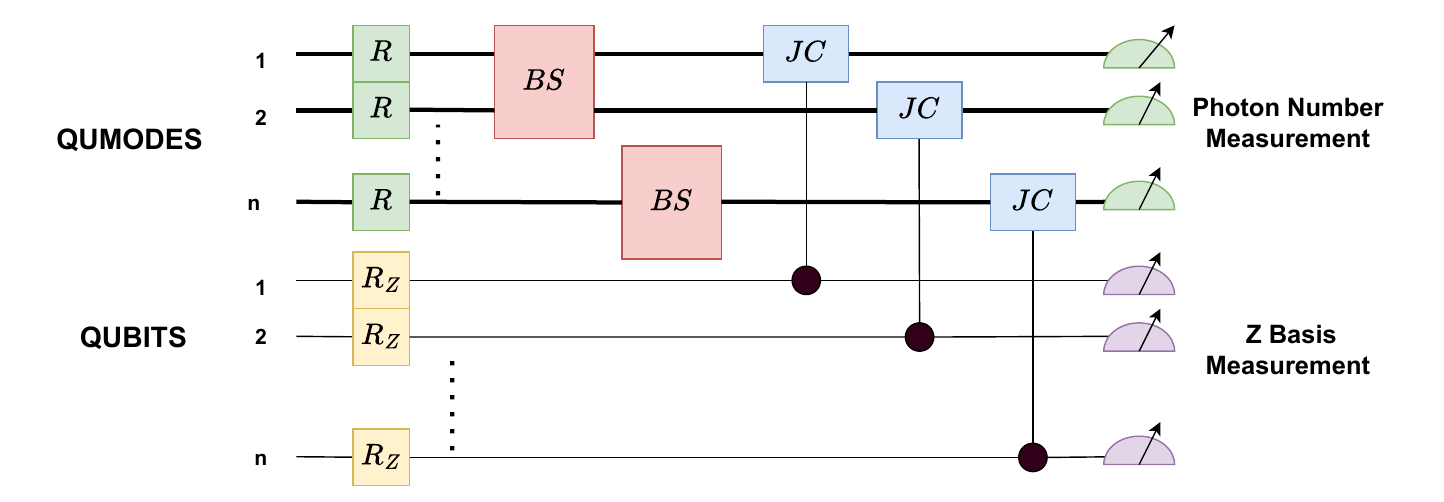}
  \caption{Circuit representation for JCH Simulation in one Trotter timestep. $R_z$ is the single qubit rotation gate, $R$ is the phase space rotation gate for a single qumode, $BS$ is the beamsplitter gate between two qumodes, and $JC$ is the Jaynes-Cummings gate between a qumode and a qubit. In each Trotter timestep, we append this subsequence.}
  \label{fig:jch_circuit}
\end{figure}

The Jaynes–Cummings–Hubbard (JCH) model describes an array of coupled cavities, each containing a two-level system (TLS), with photon hopping between neighboring sites. The Hamiltonian is

\begin{align}
H = 
& \sum_{n=1}^{N} \omega_c\, a_n^\dagger a_n 
+ \sum_{n=1}^{N} \omega_{\mathrm{tls}}\, \sigma_n^+ \sigma_n^- \notag \\
& + \kappa \sum_{n=1}^{N} \left( a_{n+1}^\dagger a_n + a_n^\dagger a_{n+1} \right) 
+ \eta \sum_{n=1}^{N} \left( a_n \sigma_n^+ + a_n^\dagger \sigma_n^- \right).
\end{align}

Because the terms in $H$ do not commute, the time-evolution operator $U(t) = e^{-iHt}$ is approximated using Trotterization, decomposing the evolution into a product of exponentials of each term over small time steps. In a hybrid CV–DV platform, each component is naturally implemented as:

\begin{itemize}
    \item \textbf{Photon energy term} $\sum_n \omega_c a_n^\dagger a_n$: implemented using phase rotation gates on each qumode.
    \item \textbf{TLS energy term} $\sum_n \omega_{\mathrm{tls}} \sigma_n^+ \sigma_n^-$: implemented with $R_z$ rotations on each qubit.
    \item \textbf{Photon hopping term} $\kappa \sum_n (a_{n+1}^\dagger a_n + a_n^\dagger a_{n+1})$: implemented using beamsplitter gates acting on neighboring qumodes.
    \item \textbf{Cavity–TLS interaction term} $\eta \sum_n (a_n \sigma_n^+ + a_n^\dagger \sigma_n^-)$: implemented with Jaynes–Cummings gates.
\end{itemize}

This decomposition highlights the advantage of hybrid CV–DV architectures, which can directly implement bosonic operations without requiring large qubit overhead for Fock space truncation.
Fig~\ref{fig:jch_circuit} shows the CV-DV circuit for one Trotter timestep.

\subsection{Shor's algorithm}
Shor’s algorithm provides an exponential speedup for integer factorization by reducing the problem to finding the period \( r \) of the modular function \( f(x) = a^x \bmod N \), where \( a \) is chosen coprime to \( N \). Once \( r \) is found using QFT, classical post-processing efficiently yields the factors of \( N \). The CV–DV implementation proposed in \cite{cvdv-shors} performs the entire period-finding procedure using only three qumodes and one qubit, independent of \( N \). The protocol exploits the fact that position and momentum operators are Fourier transforms of each other, eliminating the need for an explicit QFT. Instead, all operations are performed in the position basis, and measuring the first qumode in the momentum basis at the end effectively performs the Fourier transform.

\begin{figure}
\centering
  \includegraphics[scale=0.2]{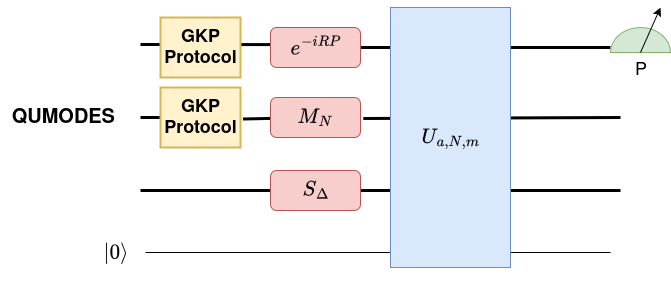}
  \caption{High-level circuit representation of the CV-DV Shor's algorithm}
  \label{fig:shors_circuit}
\end{figure}
Fig~\ref{fig:shors_circuit} provides a high-level description of the circuit for this algorithm. Two of the qumodes are initialized in approximate GKP states, while the third qumode begins in a squeezed vacuum state. The gates \( e^{-iRP} \) and \( M_N \) correspond to translation and multiplication, respectively, and can be implemented using displacement and single-mode squeezing gates. The core of the algorithm is the unitary \( U_{a,N,m} \), which applies the modular exponentiation \( x \mapsto a^x \bmod N \) in the CV space. This unitary is constructed using repeated applications of the controlled additions and multiplications in position space, combined with its inverse operations. The parameter \( m \) controls the number of repetitions, while \( R \) and \( \Delta \) determine the scaling and squeezing of the position eigenstates. After the modular exponentiation is applied, the first qumode is measured in the momentum basis, yielding an outcome from which the period \( r \) can be deduced with high probability. This CV–DV approach achieves \( \mathcal{O}(n^2) \)\cite{cvdv-shors} gate complexity for factoring an \( n \)-bit integer and provides explicit parameter ranges for \( R \), \( m \), and \( \Delta \) to ensure high fidelity of the result.

\section{Methodology}
HyQBench was built using both Bosonic Qiskit\cite{bosonic-qiskit} and QuTiP. Our Bosonic Qiskit code implements the gate-level circuit, while the QuTiP implementation is based on directly applying the unitary matrices to the state of the system, which is used to verify the correctness of the Bosonic Qiskit circuits. 

Bosonic Qiskit is an extension of Qiskit built specifically for Hybrid CV-DV systems. Bosonic Qiskit has two main register types: \textit{QubitRegister} and \textit{QumodeRegister}. The QubitRegister is used to represent the DV system or the qubits. The QumodeRegister models the CV system or the qumodes. The qumodes are built using n qubits, thereby having a cutoff of $2^n$.
Bosonic Qiskit uses the QASM simulator in the backend to simulate these circuits which has a limit of 32 qubits. Another advantage of using Bosonic Qiskit is the feature of adding custom CV-DV gates. Since it is built on top of Qiskit, we can use the UnitaryGate function to create custom unitaries for certain benchmarks. In Bosonic Qiskit, we have an all to all-to-all connectivity between the qubits and qumodes. This means that each qubit and qumode is connected to all other qubits and qumodes.

In this work, we primarily use the gates available in Bosonic Qiskit to implement the benchmark circuits. Such gates include: Displacement, Conditional Displacement, Phase Space Rotation, Jaynes-Cummings, Single Mode Squeezing, and Beamsplitter. We have also added custom gates for the CV QAOA and State Transfer circuits. The code for the benchmarking suite can be found \href{https://github.com/shubdeepmohapatra01/HyQBench/tree/main}{here}.

\section{Benchmark Results}
\subsection{State Transfer Circuit}

The state transfer circuit is implemented in Bosonic Qiskit using custom $V_j$ and $W_j$ gates constructed with the \texttt{UnitaryGate} functionality. For this example, six qubits are used to represent the qumode, and the goal is to transfer this qumode state to four qubits. The spacing parameter $\Delta$ is set to 0.39, and additional basis transformation and reverse-order measurement steps are applied as described in \cite{ad/da-cvdv} to achieve accurate state transfer.

\begin{figure}
\centering
  \includegraphics[scale=0.35]{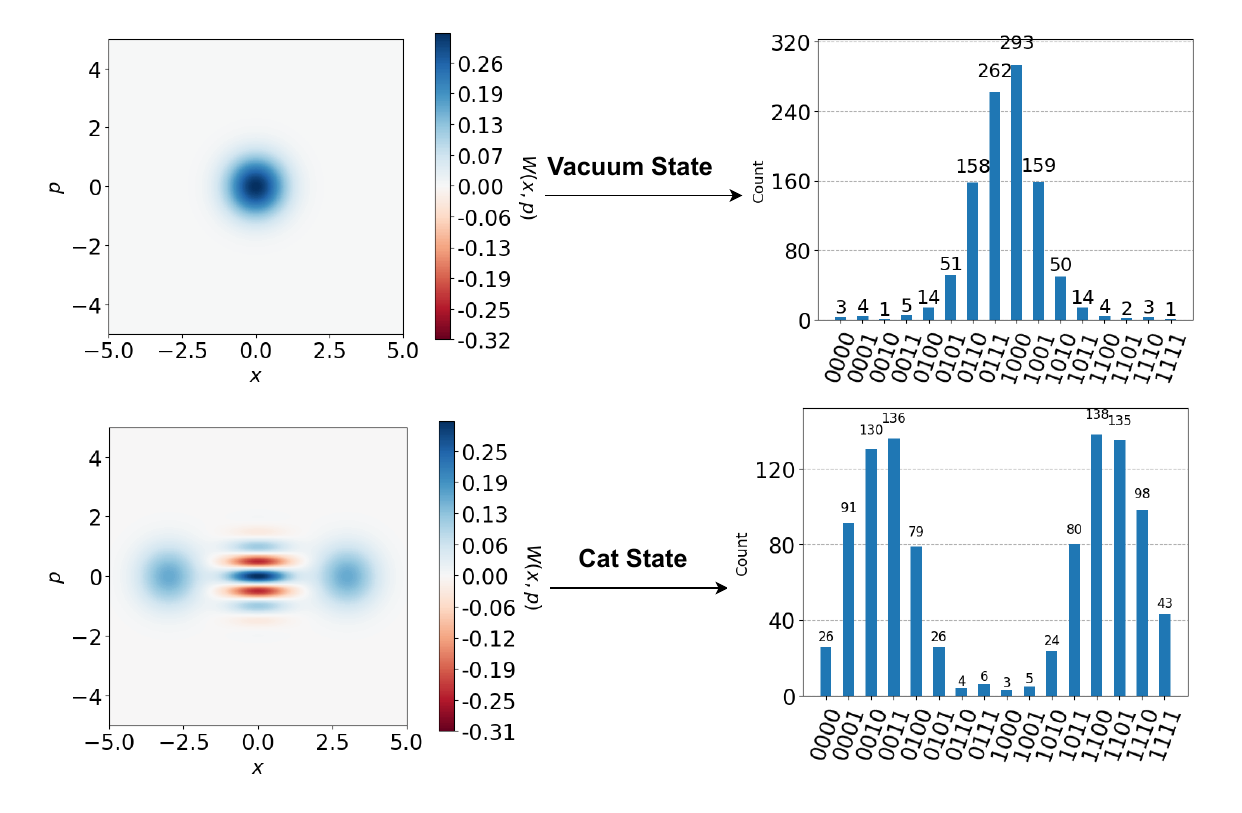}
  \caption{DV states obtained from CV-to-DV state transfer for the vacuum state and the cat state.}
  \label{fig:state_transfer2}
\end{figure}

Fig.~\ref{fig:state_transfer2} shows the DV representation of two CV states: the vacuum state and the cat state. The vacuum state, being a Gaussian Fock state centered at zero, maps to a Gaussian-like distribution in the DV basis centered around the middle bitstrings. The cat state, a superposition of two coherent states, is reflected as two distinct Gaussian peaks in the DV representation. This demonstrates that the circuit transfers these CV states to DV qubits.

\subsection{GKP State Protocol}

\begin{figure}
\centering
  \includegraphics[scale=0.35]{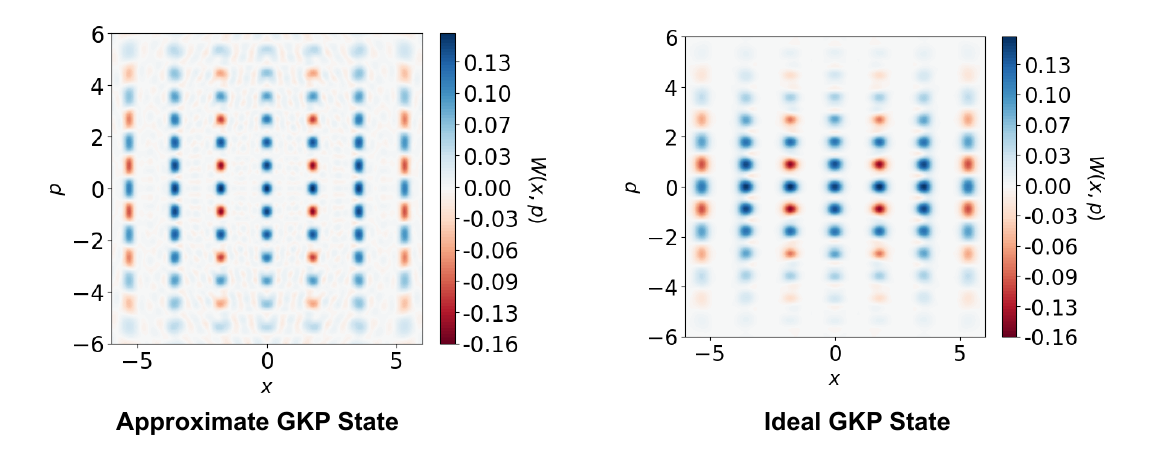}
  \caption{Approximate GKP state in Bosonic Qiskit vs.\ ideal GKP state in QuTiP.}
  \label{fig:gkp_code2}
\end{figure}

The GKP state protocol is implemented by repeatedly applying the cat state generation protocol, starting from a squeezed vacuum state to improve fidelity. Fig.~\ref{fig:gkp_code2} compares the approximate GKP state generated in Bosonic Qiskit with the ideal GKP state from QuTiP. For a cutoff of 64 (6 qubits), a squeezing parameter of 0.222, and nine repetitions of the cat state protocol, the resulting fidelity is 0.66. This fidelity result arises from the fixed value $\alpha = \sqrt{\pi}$ for GKP states, where the overlap between the displaced coherent states is non-negligible. Increasing the squeezing parameter can also increase fidelity, however high squeezing operation is expensive to implement. \cite{cvdv-qsp} provides another protocol for generating higher-fidelity GKP states by using deeper circuits.

\subsection{CV-DV QFT}

The CV-DV QFT algorithm transfers the state of $n$ qubits to a qumode, applies a Fourier transform on the qumode, and transfers the state back to qubits. The circuit includes basis transformations, displacement gates with optimized spacing parameters, and the addition of ancilla and append qubits to improve periodicity and fidelity. For a 2-qubit input state $\ket{00}$, we use one ancilla qubit, two append qubits, a cutoff of 16 (4 qubits for the qumode), and spacing parameters $\Delta = 2.33$ and $0.29$ for the displacement gate and state transfer protocols, respectively.  

\begin{figure}
\centering
  \includegraphics[scale=0.25]{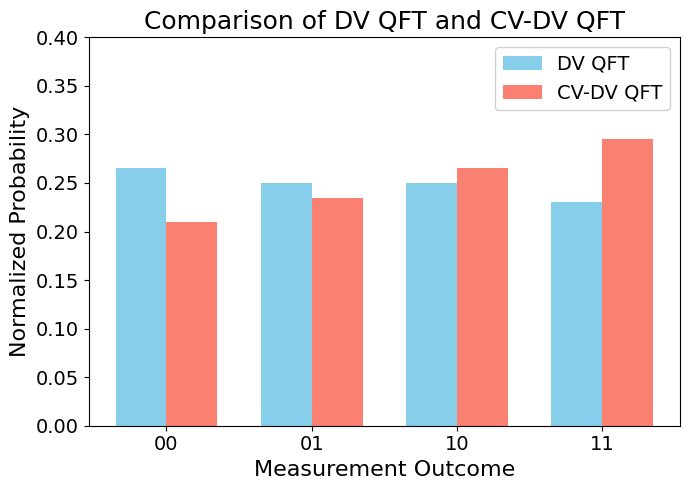}
  \caption{State comparison of DV QFT vs.\ CV-DV QFT.}
  \label{fig:cv_qft}
\end{figure}

Fig.~\ref{fig:cv_qft} compares the output of the CV-DV QFT with the standard DV QFT. The resulting fidelity is 0.94, and adding more ancilla qubits further increases the fidelity.  

\subsection{CV-QAOA}

The CV-QAOA benchmark demonstrates optimization in a purely CV circuit using alternating applications of the cost and mixer unitaries $e^{-i \eta_j \hat{H}_C}$ and $e^{-i \gamma_j \hat{H}_M}$, where $\hat{H}_C = (\hat{x}-3)^2$ and $\hat{H}_M = \frac{\hat{p}^2}{2}$. The circuit, initialized in a squeezed vacuum state, was optimized using a depth of 5 and the BFGS classical optimizer.  

\begin{figure}
\centering
  \includegraphics[scale=0.25]{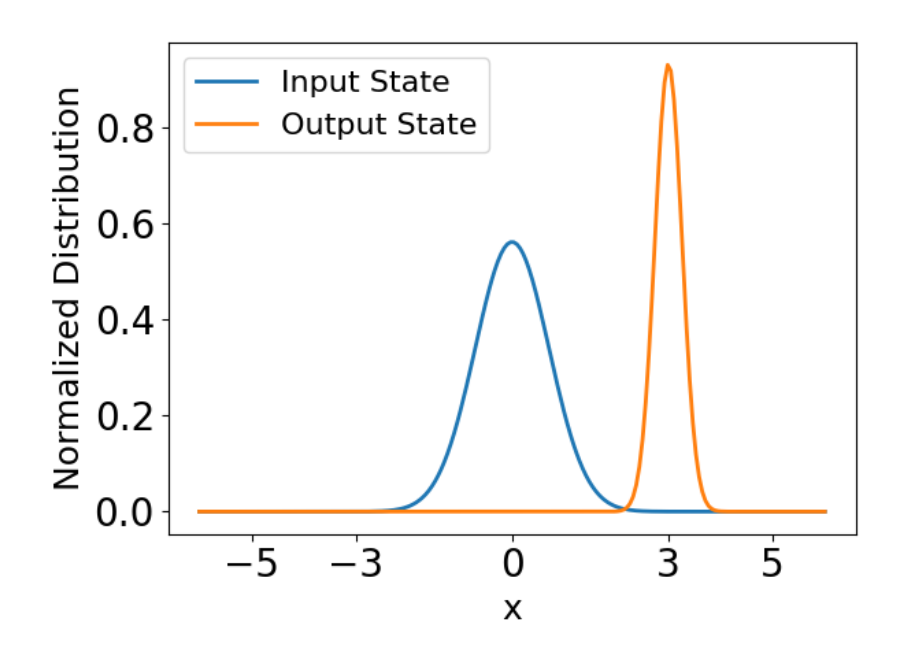}
  \caption{Initial and final state distributions along the position axis.}
  \label{fig:cv_qaoa1}
\end{figure}

Fig.~\ref{fig:cv_qaoa1} shows that the final state's position distribution is centered around $x=3$, corresponding to the minimum of the target function. While the example uses a simple quadratic function, the method can be extended to more complex cost functions with appropriate gate decompositions.  

For polynomials with a degree higher than two in $\hat{x}$ or $\hat{p}$, the cost unitary will be non-Gaussian in nature. As non-Gaussian gates in the CV-only setting are more expensive, it would be beneficial to implement them as hybrid CV-DV gates. 
However, the decomposition of the cost unitary into hybrid gates is not trivial.

\subsection{CV-DV VQE}

\begin{figure}
\centering
  \includegraphics[scale=0.3]{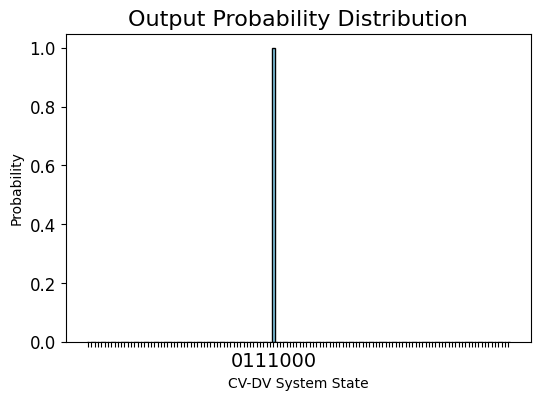}
  \caption{Probability distribution of the final qubit–qumode state. Note that here the state 0111000 is the binary representation of the combined qubit state, qumode 1 photon number, and qumode 2 photon number. Grouping them into item and auxiliary variables gives the state (0$|$1,1,1$|$0,0,0).}
  \label{fig:vqe1}
\end{figure}

We implemented a CV-DV VQE circuit in Bosonic Qiskit to solve a binary knapsack problem with 4 items having values \([1,4,5,10]\), weights \([2.5,1,2,3]\), and knapsack capacity \(W=7\). The qubit and first qumode (cutoff $2^3$) represented the item variables, while the second qumode (cutoff $2^3$) encoded the auxiliary variables. The circuit, with depth 5 and the classical BFGS optimizer, converged to the optimal solution $(x_1,x_2,x_3,x_4) = (0,1,1,1)$ after 117 iterations, achieving a total value of 19 and weight 6 (Fig.~\ref{fig:vqe1}). This result shows that arbitrary binary knapsack problems can be solved using just two qumodes and one qubit by scaling the Fock cutoff, though higher cutoffs increase susceptibility to photon loss errors.

\subsection{JCH Hamiltonian Simulation}

We simulated the 1D Jaynes–Cummings–Hubbard (JCH) model in Bosonic Qiskit using Trotterization with a time step of 0.1 for 50 steps (total time 5 units). The system consisted of three cavities, each coupled to a TLS, with photon hopping allowed between neighboring cavities. The parameters (in natural units) were set to $\omega_c = \omega_{\mathrm{tls}} = 4\pi$, $\kappa = 1$ (strong hopping), $\eta = 0.5$ (moderate coupling), and a cutoff of 8 (3 qubits per qumode). The leftmost cavity was initialized with two photons.  

\begin{figure}
\centering
  \includegraphics[scale=0.25]{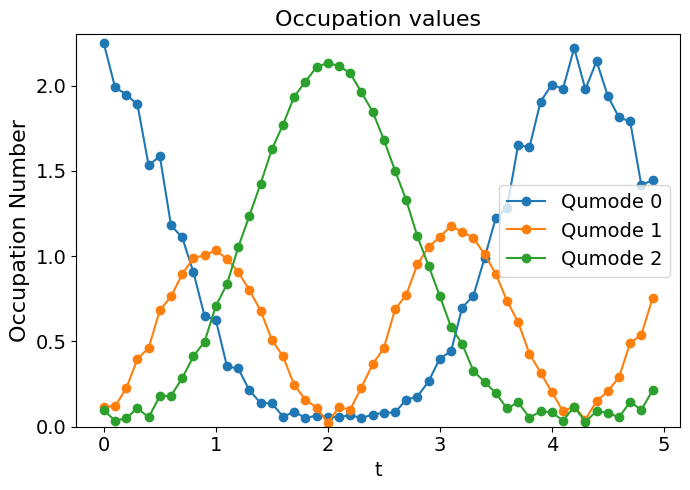}
  \caption{Photon number occupation for each qumode in the JCH Hamiltonian simulation.}
  \label{fig:jch2}
\end{figure}

Fig.~\ref{fig:jch2} shows the photon number evolution in each cavity. Qumodes 0 and 2 (edges) oscillate between 0 and 2 photons, while Qumode 1 (middle) oscillates between 0 and 1 photon. The strong hopping $\kappa=1$ causes photons to hop rapidly across cavities, and the total photon number remains conserved at 2 throughout the simulation.

\subsection{Shor's Algorithm}

We implemented Shor's algorithm in Bosonic Qiskit using three qumodes and one qubit, following the CV-DV approach of \cite{cvdv-shors}. The key advantage is that the Fourier transform is performed implicitly by measuring in the momentum basis, removing the need for an explicit QFT. The two qumodes were initialized with approximate GKP states generated using the previous protocol, with squeezing parameter 1.202 and nine repetitions of the cat-state protocol. Due to the QASM simulator limit of 32 qubits, we used a cutoff of $2^{10}$ for the qumodes, which allowed factoring numbers up to 1024.  

\begin{figure}
\centering
  \includegraphics[scale=0.3]{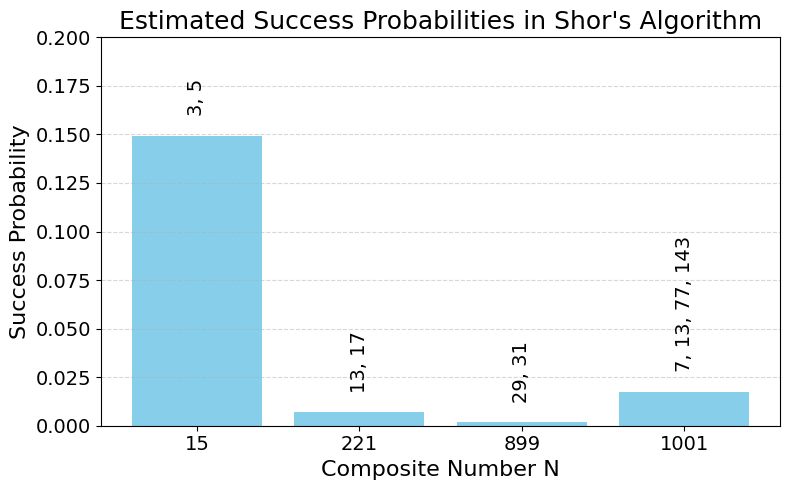}  
  \caption{Success probability of finding factors for different values of $N$ using CV-DV Shor's algorithm.}
  \label{fig:shors_result}
\end{figure}

Fig.~\ref{fig:shors_result} shows the success probability for factoring $N=${15, 221, 899, 1001} over 5 trials each, with random coprime choices of $a$. The unitary $U_{a,N,m}$ was implemented with $m=2$. The overall success probability was limited by the use of approximate GKP states (fidelity $\approx 0.66$), but the algorithm successfully produced nontrivial factors for all tested values of $N$.

\section{Benchmark Characterization}

\subsection{ Characterization Metrics}
To characterize the hybrid CV--DV circuits, we propose the following two categories of metrics.

\paragraph{General Structure Features:}  
These include the number of qubits, qumodes, qubit/qumode gates, hybrid gates, and the circuit depth.

\paragraph{CV-DV Features:}  
We propose these circuit features that are particularly relevant to the Hybrid CV-DV Systems:
\begin{itemize}
    \item \textbf{Wigner Negativity:} The Wigner function is a quasiprobability distribution that can take negative values, which serve as signatures of non-classicality. Larger magnitude and volume of these negative regions indicate stronger quantum behavior and correspondingly higher difficulty for classical simulation\cite{wigner-negativity}. Wigner negativity is quantified by integrating the absolute value of the negative part of the Wigner function over phase space. Its minimum value is 0, while there is no upper bound, with higher values reflecting increased classical complexity.
    
    \item \textbf{Truncation Cost:} For classical simulation, the infinite-dimensional CV system is truncated at a cutoff dimension $N_{\text{cutoff}}$ which introduces error if significant probability mass lies in the higher the truncated region. To quantify this, we define the truncation cost as:
    \begin{equation}
        \mathcal{C}_{\text{trunc}} = \sum_{n = N_{\text{cutoff}} - k}^{N_{\text{cutoff}} - 1} \langle n | \rho | n \rangle.
        \label{eq:truncation_cost}
    \end{equation}
    It captures the probability of occupation in the highest $k$ Fock levels. A high truncation cost suggests that a large cutoff is necessary to reduce the classical simulation error.

    \item \textbf{Energy:} The energy of the circuit at any phase of its execution is computed using:
    \begin{equation}
        \mathcal{E}_{\text{avg}} = \sum_{j=1}^{M} \langle \hat{n}_j \rangle + \sum_{k=1}^{Q} \langle \sigma_z^{(k)} \rangle.
        \label{eq:final_state_average_energy}
    \end{equation}
    Here, $M$ is the number of qumodes and $Q$ is the number of qubits. The first term accounts for photon number expectation in each mode, and the second term for spin-based energy in the DV subsystem. We assume natural units throughout.
\end{itemize}
The selected CV-DV metrics are intended to capture both computational cost and nonclassical resource generation in hybrid CV–DV circuits. Wigner negativity serves as a proxy for non-Gaussian resource content and simulation hardness, while truncation cost reflects effective Hilbert-space dimension and memory/runtime requirements for classical simulation.

We calculate the \textbf{maximum} value of these metrics observed during the execution of the circuit. This is done by tracking the state of the circuit after each gate operation and using that to calculate the metric values. For each CV-DV specific metric, the values are normalized by dividing by the maximum observed value across all benchmarks, so that each metric ranges between 0 and 1. 

\subsection{Characterization Results}

\begin{table*}[h!]
\centering
\caption{Metrics showing the circuit features and the CV-DV specific features.}
\label{tab:metrics}

\resizebox{\linewidth}{!}{%
\begin{tabular}{|l|c|c|c|c|c|c|c|c|c|l|}
\hline
\multirow{2}{*}{Benchmark} & \multicolumn{6}{c|}{Circuit Features} & \multicolumn{3}{c}{CV-DV Specific Features} \\
\cline{2-10}
 & Qubits & Qumodes & Qubit Gates & Qumode Gates & Hybrid Gates & Circuit Depth & Max Energy & Max Wigner Negativity & Max Truncation Cost \\
\hline
StateTransferCVtoDV & 4 & 1 & 9 & 0 & 8 & 12 & 0.12 & 0.14 & 0.24 \\
Cat State           & 1 & 1 & 6 & 0 & 2 & 8 & 0.15 & 0.09 & 0.19 \\
GKP State           & 1 & 1 & 48 & 1 & 16 & 64 & 0.23 & 0.30 & 0.90 \\
QFT Circuit         & 5 & 1 & 23 & 3 & 20 & 29 & 0.19 & 0.19 & 0.58 \\
CV-DV VQE           & 1 & 2 & 10 & 0 & 10 & 20 & 0.09 & 0.13 & 1.00 \\
CV QAOA             & 0 & 1 & 0 & 11 & 0 & 11 & 0.28 & 1.00 & 0.26 \\
JCH N=3 (per Trotter step)             & 3 & 3 & 3 & 5 & 3 & 4 & 0.08 & 0.05 & 0.00 \\
Shor's Circuit      & 1 & 3 & 128 & 32 & 80 & 209 & 1.00 & 0.59 & 0.06 \\
\hline
\end{tabular}%
}
\end{table*}

Table~\ref{tab:metrics} shows the metric values for the CV-DV specific features and the circuit features. Taken together, these benchmarks illustrate two distinct behaviors. 
First, circuits such as JCH simulation, state transfer protocol, and the Cat State circuit exhibit low Wigner negativity, energy, and low truncation cost. These features explain why their dynamics are more suitable for classical simulation, as the states remain close to Gaussian and occupy mostly the lower Fock levels. Second, algorithms such as CV-QAOA and Shor’s circuit display significantly higher Wigner negativity, reflecting strong non-classicality, which makes them more difficult to simulate on classical devices. CV-DV VQE provides a complementary case, where the Wigner negativity is low but the truncation cost is high, indicating that simulation hardness can also arise from choosing an unsuitable cutoff value for the qumodes.

\begin{figure}[t]
\centering
\includegraphics[width=\linewidth]{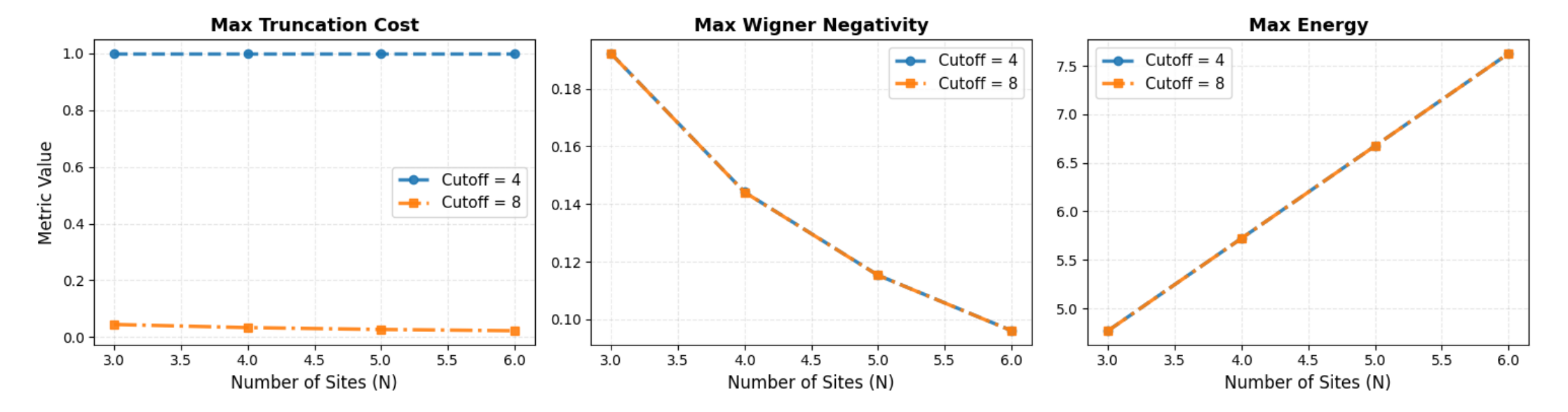}
\caption{Metric comparison for JCH Simulation for different problem sizes and Fock-level cutoffs.
}
\label{fig:jch_usecase}
\end{figure}
To illustrate the practical meaning of our metrics, we analyze how the JCH simulation behaves as we vary both the number of sites ($N$), i.e., the number of qumodes, and the cutoff ($2^n$, where n is the number of qubits required to represent each qumode). Fig~\ref{fig:jch_usecase} compares the truncation cost, Wigner negativity, and energy of the JCH circuit for cutoffs $= 4$ and $8$, with all other values remaining the same as used in Section~V.F. As $N$ increases, the energy scales approximately linearly, while the Wigner negativity gradually decreases, implying that larger systems become more efficient to simulate classically. Increasing the cutoff from $4$ to $8$ sharply lowers the truncation cost from $\sim1$ to $\sim0.02$. This demonstrates a key tradeoff: higher cutoff yields lower the chances of truncation errors at the expense of exponentially higher simulation cost. However, increasing the cutoff, does not affect the Wigner negativity and the Energy, since these metrics do not depend on the Fock cutoff but rather on the state of the qumodes.

\section{Benchmark Selection}
The benchmarks in HyQBench were selected to provide hierarchical coverage across three abstraction levels—primitives (state preparation, state transfer), algorithms (QFT, VQE, QAOA), and applications (JCH simulation, Shor's algorithm). HyQBench is designed as an extensible open-source framework; new benchmarks can be contributed by implementing circuits in Bosonic Qiskit.

\begin{figure}
\centering
  \includegraphics[scale=0.25]{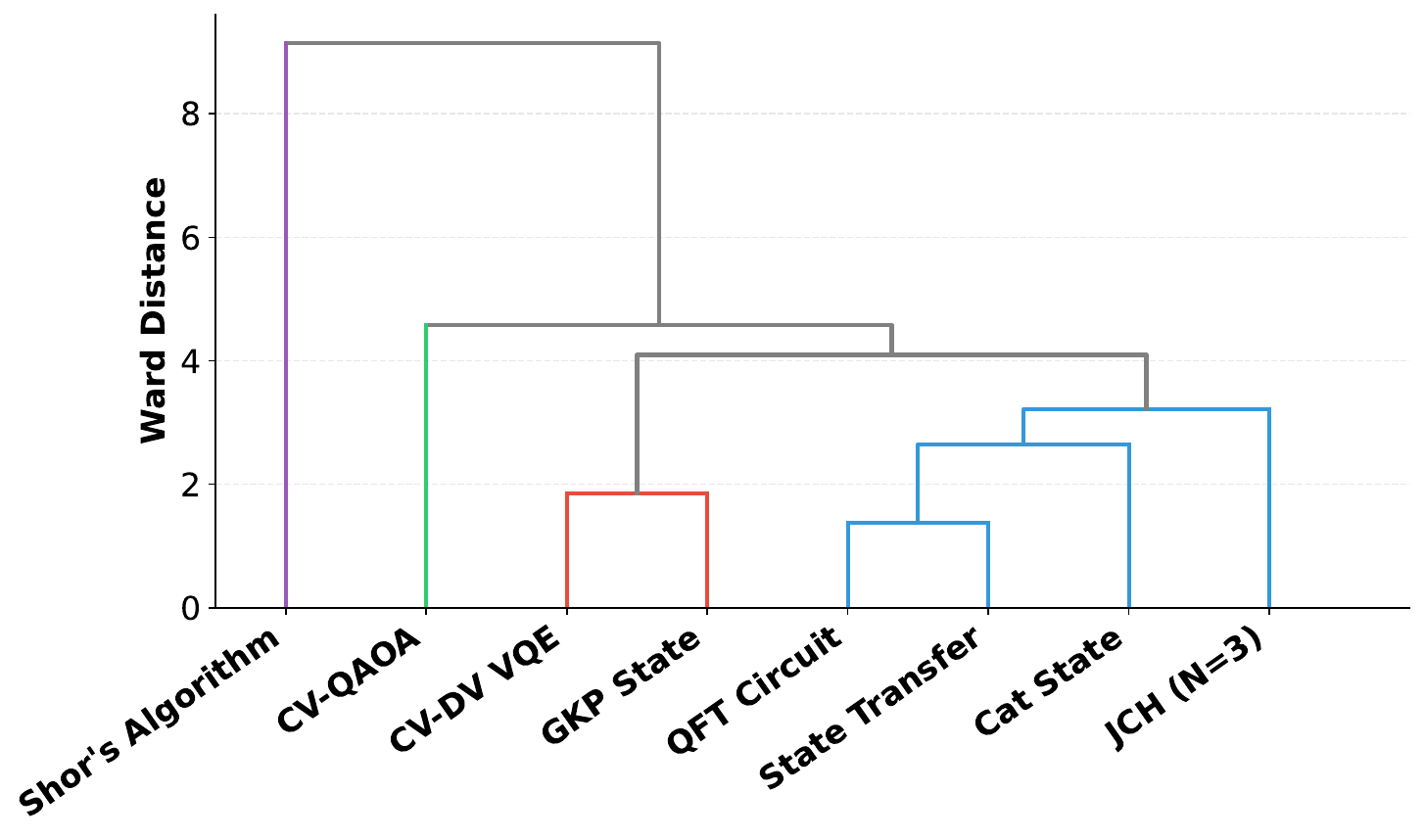}
  \caption{Dendrogram showing the circuit coverage}
  \label{fig:benchmark_dendrogram}
\end{figure}

Fig.~\ref{fig:benchmark_dendrogram} presents hierarchical clustering of the eight HyQBench benchmarks using Ward's method on standardized circuit features. Circuit features were standardized via z-score normalization across benchmarks before clustering. Four distinct clusters emerge.

Cluster 1 (High Truncation Cost) contains GKP State and CV-DV VQE, both characterized by truncation cost z-scores exceeding +1.5, indicating large Fock cutoffs are required for accurate simulation. Cluster 2 (Primitives and Algorithms) groups State Transfer, Cat State, QFT Circuit, and JCH Simulation—benchmarks with moderate complexity and no extreme feature values. Cluster 3 (High Wigner Negativity) isolates CV-QAOA, which exhibits a Wigner negativity z-score of +2.26, reflecting strongly non-classical state generation. Cluster 4 (High Complexity) contains only Shor's Algorithm, distinguished by extreme values across all complexity metrics.

This clustering indicates that HyQBench benchmarks occupy distinct regions of the feature space, supporting broad coverage of hybrid CV–DV workload characteristics.

\section{Advantage of Hybrid CV-DV over CV-only or DV-only Systems}
\begin{table}[h]
\centering
\caption{Resource requirement comparison among DV-only, CV-only, and CV--DV systems.}
\renewcommand{\arraystretch}{1}
\resizebox{\columnwidth}{!}{%
\begin{tabular}{|l|c|c|c|}
\hline
\textbf{Benchmark} & \textbf{DV-only} & \textbf{CV-only}  & \textbf{CV--DV} \\
\hline
State Transfer & 
\begin{tabular}[c]{@{}c@{}}N/A\end{tabular} & 
\begin{tabular}[c]{@{}c@{}}N/A\end{tabular} & 
\begin{tabular}[c]{@{}c@{}}Qubits: $n$\\Qumodes: $1$\\Gates: $O(n)$ \cite{ad/da-cvdv}\\\end{tabular} \\
\hline
Cat State  & 
\begin{tabular}[c]{@{}c@{}}N/A\end{tabular} & 
\begin{tabular}[c]{@{}c@{}}N/A\\\end{tabular} & 
\begin{tabular}[c]{@{}c@{}}Qubits: $1$\\Qumodes: $1$\\Gates: $O(1) $\cite{bosonic-isa}\\\end{tabular} \\
\hline
GKP State & 
\begin{tabular}[c]{@{}c@{}}N/A\end{tabular} & 
\begin{tabular}[c]{@{}c@{}}Qumodes: $1$\\Gates: $O(1)$ \cite{cv-gkp}\\\end{tabular} & 
\begin{tabular}[c]{@{}c@{}}Qubits: $1$\\Qumodes: $1$\\Gates: $O(1)$\\\end{tabular} \\
\hline
QFT & 
\begin{tabular}[c]{@{}c@{}}Qubits: $n$\\Gates: $O(n\log (\frac{n}{\epsilon}))$\cite{qft-resource}\end{tabular} & 
\begin{tabular}[c]{@{}c@{}}N/A\end{tabular} & 
\begin{tabular}[c]{@{}c@{}}Qubits: $n+a$\\Qumodes: $1$\\Gates: $O(n+a)$\cite{ad/da-cvdv}\\\end{tabular} \\
\hline
VQE (per layer) & 
\begin{tabular}[c]{@{}c@{}}Qubits: $n$\\Gates: $O(n)$\cite{vqe-resource}\end{tabular} & 
\begin{tabular}[c]{@{}c@{}}N/A\end{tabular} & 
\begin{tabular}[c]{@{}c@{}}Qubits: $1$\\Qumodes: $2$\\Gates: $4$\cite{cvdv-vqe}\\\end{tabular} \\
\hline
QAOA (per layer) & 
\begin{tabular}[c]{@{}c@{}}Qubits: $n$\\Gates: $O(n)$\cite{dv-qaoa}\end{tabular} & 
\begin{tabular}[c]{@{}c@{}}Qumodes: $1$\\ Unitary Operations: $O(1)$\cite{cv-qaoa}\end{tabular} & 
\begin{tabular}[c]{@{}c@{}}N/A\end{tabular} \\
\hline
JCH (cutoff $m$) & 
\begin{tabular}[c]{@{}c@{}}Qubits: $n+n\log m$\\Gates: $O(n m\log m)$\cite{jch-resource}\end{tabular} & 
\begin{tabular}[c]{@{}c@{}}N/A\end{tabular} & 
\begin{tabular}[c]{@{}c@{}}Qubits: $n$\\Qumodes: $n$\\Gates: $O(n)$\\\end{tabular} \\
\hline
Shor's Factoring & 
\begin{tabular}[c]{@{}c@{}}Qubits: $2n+2$\\Gates: $O(n^3 \log n)$\cite{shors-resource}\end{tabular} & 
\begin{tabular}[c]{@{}c@{}}N/A\end{tabular} & 
\begin{tabular}[c]{@{}c@{}}Qubits: $1$\\Qumodes: $3$\\Gates: $O(n^2)$\cite{cvdv-shors}\\\end{tabular} \\
\hline
\end{tabular}%
}
\label{tab:resources}
\end{table}

Compared to DV-only or CV-only quantum systems, hybrid CV-DV systems offer advantages as discussed in Section II.C. Here, we provide analytical results to quantify such advantages in resource requirements for each of the benchmarks, as presented in 
Table~\ref{tab:resources}. 

Hybrid CV–DV architectures provide resource advantages across several benchmark tasks. In many cases, a single qumode offers a large, native Hilbert space that reduces the number of qubits and gate operations required. Tasks such as state transfer, cat-state preparation, GKP generation, and hybrid QFT benefit from constant or linear gate complexity by leveraging hybrid Cv-DV operations. Variational algorithms such as VQE and QAOA also require significantly fewer qubits, and bosonic Hamiltonians like the Jaynes–Cummings–Hubbard model become easier to simulate due to the native CV dynamics. Shor’s algorithm in hybrid form uses a fixed number of qumodes and achieves lower gate cost.

These analytical results on the benchmarks showcase the advantages of hybrid CV-DV systems, including the high dimensionality of a qumode, the ease of simulating bosonic terms in a Hamiltonian and the advantage of directly using subroutines like QFT as evidenced in Shor's algorithm.

\section{Noise Analysis}
Photon loss is the dominant decoherence mechanism in continuous-variable (CV) systems, arising from scattering and absorption at cavity boundaries. In Bosonic Qiskit, photon loss is modeled using the Kraus operators of the photon-loss channel, parameterized by the annihilation operator $\hat{a}$ and decay rate $\kappa$, the inverse of the cavity lifetime. Typical cavity lifetimes range from $100~\mu s$ to $10~ms$,\cite{cavity-lifetime1,cavity-lifetime3}, with some exceeding these limits,\cite{cavity-lifetime2}. We assume a cavity lifetime of $1~ms$, giving $\kappa = 1000~\text{s}^{-1}$.
Gate durations depend inversely on the qubit–cavity coupling strength $\chi$, which typically lies in the 1–3 MHz range; here we set $\chi = 1$~MHz and compute hybrid gate times accordingly, $\sim \frac{\text{gate parameter}}{\chi}$ (eg, $\frac{\alpha}{\chi}$ for CD gate, $\frac{\theta}{\chi}$ for CR gate). We assume the DV subsystem to have damping ($T1 = 30 \mu s$) and dephasing noise ($T2 = 65 \mu s$)\cite{cavity-lifetime3}. 

\begin{table}[h!]
\centering
\caption{Benchmark Fidelity in Noisy and Noise-Free settings.}
\begin{tabular}{|l|c|c|c|}
\hline
\textbf{Benchmark} & \textbf{Total Circuit Duration} & \textbf{Fidelity} \\
\hline
State Transfer (DV to CV) & $4.7\,\mu\text{s}$ & 0.99 \\
Cat State & $0.8\,\mu\text{s}$ & 0.99 \\
GKP State & $5.6\,\mu\text{s}$ & 0.97 \\
QFT Circuit & $9.4\,\mu\text{s}$ & 0.99 \\
CV-DV VQE & $5.2\,\mu\text{s}$ & 0.91 \\
JCH N=3 (10 Trotter steps) & $20.6\,\mu\text{s}$ & 0.92 \\
JCH N=3 (50 Trotter steps) & $0.1\,\text{ms}$ & 0.03 \\
Shor's Circuit & $0.9\,\text{ms}$ & 0.1 \\
\hline
\end{tabular}
\label{tab:fidelity}
\end{table}

Table~\ref{tab:fidelity} shows the Uhlmann Fidelities for all benchmark circuits except QFT in noisy settings (i.e., photon loss noise channel) compared to noise-free ones. 
We have not included CV-QAOA as the gate decomposition for the circuit is still an open problem. 
For the QFT benchmark, fidelity is computed on the DV subsystem alone with both CV and DV noise being considered. The reason is that the goal of the circuit is to implement the QFT on the DV qubits using CV modes, and the final measurement is performed solely on the DV qubits.
Overall, the circuits with short durations (Cat, GKP, QFT, VQE, State Transfer, JCH for 10 Trotter steps) maintain high fidelity because their operation times are well below the $1ms$ cavity lifetime. In contrast, the JCH (50 Trotter steps) and Shor’s circuits suffer substantial degradation due to their much longer execution times, which accumulate significant photon loss. 
At present, the Bosonic Qiskit simulation stack used in this work supports photon-loss channels for the bosonic mode, which we use as the primary noise model for the initial benchmark noise analysis. Joint qubit–qumode noise fitting is currently in progress and will be incorporated in future benchmark releases.

\section{Real Machine Results}

We have run the CAT State Benchmark on QSCOUT's trapped-ion platform. While the device is still undergoing active calibration of its CV–DV gate set, these experiments already demonstrate the feasibility of hybrid CV–DV systems and highlight the need for a dedicated benchmarking suite.

The CAT State Benchmark is currently being used to calibrate the displacement amplitudes of QSCOUT’s Conditional Displacement (CD) gate. In this experiment, the CAT state is reconstructed through measurements of the \textbf{characteristic function}\cite{cf-fidelity}, defined for a density operator $\rho$ as
\begin{equation}
\chi(\beta)
=
\mathrm{Tr}\left[\rho D(\beta)\right],
\qquad
\beta \in \mathbb{C}, 
\label{eq:cf}
\end{equation}
where $D(\beta) = \exp\!\left(\beta a^\dagger - \beta^\ast a\right)$ is the displacement operator.
$\chi(\beta)$ is the Fourier transform of the Wigner quasiprobability distribution.

\begin{figure}[t]
\centering
\includegraphics[width=0.8\linewidth]{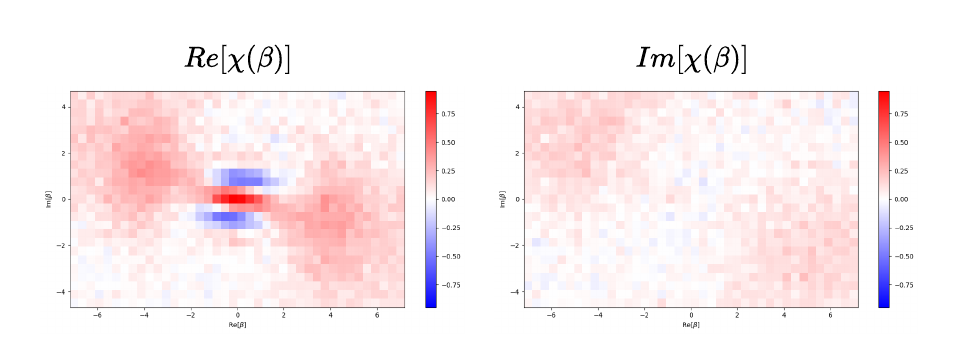}
\caption{Real and imaginary parts of the characteristic function for the CAT state measured on QSCOUT.}
\label{fig:cat_qscout}
\end{figure}

Fig.~\ref{fig:cat_qscout} shows the real and imaginary components of the measured characteristic function using QSCOUT's device. The horizontal and vertical axes correspond to the real and imaginary parts of the displacement parameter $\beta$ in Eq.~\ref{eq:cf}. The fidelity between two states $\rho$ and $\sigma$ can be computed directly from their characteristic functions \cite{cf-fidelity}:
\begin{equation}
F(\rho,\sigma)
=
\frac{1}{\pi}
\int_{\mathbb{C}}
d^{2}\beta
\chi_\rho(\beta)
\chi_\sigma^\ast(\beta).
\label{eq:FidChi}
\end{equation}
Using this formula, the fidelity of the CAT state generated using the QSCOUT device is found to be $0.71$. The observed infidelity is dominated by imperfect gate calibration in the CD gate amplitudes instead of other error sources like dephasing and heating errors. This further motivates the use of our benchmarks to calibrate the CV-DV gates in CV-DV systems.

\section{Conclusion}

In this work, we develop a framework for benchmarking hybrid continuous-variable (CV)–discrete-variable (DV) quantum circuits. By constructing and analyzing benchmark circuits such as cat state preparation, CV-DV quantum Fourier transform, and variational quantum algorithms, Hamiltonian Simulation, we demonstrate the versatility and computational potential of hybrid architectures.

Our proposed circuit features, which include both general structural features and CV-DV-specific indicators such as Wigner negativity and truncation cost, provides a systematic way to assess the complexity and scalability of hybrid circuits. These metrics help identify trade-offs between circuit expressiveness, classical simulability, and qumode resource requirements.

We also present a noise analysis and real machine experiment results for the benchmark. A more accurate hardware-specific noise modeling/analysis is left for future works.

This work lays the foundation for more structured evaluation of hybrid quantum algorithms and can serve as a stepping stone experimental implementation of hybrid architectures.

\section{Acknowledgements}

This work is supported in part by  the U.S. Department of Energy (DOE), Office of Science, Office of Advanced Scientific Computing Research (ASCR), under Award Number DE-SC0025384, DE-SC0025563, the NSF grant OSI-2410675, PHY-2325080 (with a subcontract to NC State University from Duke University), and OMA-2120757 (with a subcontract to NC State University from the University of Maryland). 

\bibliographystyle{ACM-Reference-Format}
\bibliography{references.bib}

\end{document}